\documentclass[apj]{emulateapj}
\usepackage{amssymb}
\usepackage{epsfig}
\usepackage{natbib}
\usepackage{url}
\usepackage{hyperref}
\usepackage{subfigure}
\usepackage{graphicx}
\usepackage[]{graphicx}
\usepackage{booktabs}
\usepackage{epstopdf}
\received{\today}
\accepted{}

\shorttitle{A tidal stream around the Whale galaxy}
\shortauthors{Mart\'{\i}nez-Delgado et al.}

\begin{document}

\title{Discovery of a stellar tidal stream around the Whale galaxy, NGC 4631\\}

\author{David Mart\'{\i}nez-Delgado\altaffilmark{1},
 Elena D'Onghia \altaffilmark{2,}\altaffilmark{3},
 Taylor S. Chonis \altaffilmark{4}, 
 Rachael L.  Beaton \altaffilmark{5},
 Karel Teuwen \altaffilmark{6},
 R. Jay GaBany \altaffilmark{7}, Eva K. Grebel\altaffilmark{1}, Gustavo Morales\altaffilmark{1}}

\altaffiltext{1}{Astronomisches Rechen-Institut, Zentrum f\"ur Astronomie der Universit\"at Heidelberg, 
 M{\"o}nchhofstr. 12-14, 69120 Heidelberg, Germany}
\altaffiltext{2}{Alfred P. Sloan Fellow}
\altaffiltext{3} {University of Wisconsin, 475 N. Charter St., 53706 Madison}
\altaffiltext{4} {Department of Astronomy, University of Texas at Austin, 2515 Speedway, Stop C1400, Austin,
 TX 78712, USA}
\altaffiltext{5} {Observatories of the Carnegie Institutions, Pasadena, USA}
\altaffiltext{6} {Remote Observatories Southern Alpes, Verclause, France}
\altaffiltext{7} {Black Bird Observatory II, Alder Springs, California, USA}


\begin{abstract}

We report the discovery of a giant stellar tidal stream in the halo of NGC\,4631, a nearby edge-on spiral
galaxy interacting with the spiral NGC\,4656, in deep images taken with a 40-cm aperture robotic
telescope. The stream has two components: a bridge-like feature extended between NGC\,4631 and NGC\,4656 (stream$_{SE}$) and an overdensity with extended features on the opposite side of the NGC\,4631 disk (stream$_{NW}$). Together,  these features extend more than 85 kpc and display a clear $(g-r)$ colour gradient. The orientation of stream$_{SE}$ relative to the orientations of NGC\,4631 and NGC\,4656 is not consistent with an origin from interaction between these two spirals, and is more likely debris from a satellite encounter. The stellar tidal features can be qualitatively reproduced in an $N$-body model of the tidal disruption of a single, massive dwarf satellite on a moderately eccentric orbit (e=0.6) around NGC\,4631 over $\sim$ 3.5 Gyr, with a dynamical mass ratio (m1:m2) of $\sim$ 40. Both modelling and inferences from the morphology of the streams indicate these are not associated  with the complex HI tidal features observed between both spirals, which likely originate from a more recent, gas-rich accretion event. The detailed structure of stream$_{NW}$ suggests it may contain the progenitor of the stream, in agreement with the $N$-body model. In addition, stream$_{NW}$ is roughly aligned  with two very faint dwarf spheroidal candidates. The system of dwarf galaxies and the tidal stream around NGC\,4631 can provide an additional interesting case for exploring the anisotropy distribution of satellite galaxies recently reported in Local Group spiral galaxies by means of future follow-up observations.
\end{abstract}


\keywords{galaxies: dwarf --- galaxies: evolution
      --- galaxies: photometry --- Local Group}


\section{INTRODUCTION}

Numerical cosmological models 
 built within the  $\Lambda$-Cold Dark Matter ($\Lambda$-CDM) paradigm (Bullock \& Johnson 2005; Cooper et al. 2010; Pillepich, Madau \& Mayer 2014)  predict that deep imaging of the stellar halos of nearby large spiral galaxies 
 should display a wide variety of diffuse, low-surface brightness stellar sub-structures. 
The most common features are stellar streams or shells, 
 which result from interactions between the parent galaxy and its dwarf satellite companions. 
The detection of such features as a {\it ubiquitous} feature of all galaxies, as is implied by $\Lambda$-CDM,
 remains untested observationally. This is largely due the challenge of producing images
 with sufficient surface brightness sensitivity and low background fluctuations to reliably detect such faint features.
Although more luminous examples of diffuse stellar streams and shells around massive elliptical galaxies 
 have been known for many decades (e.g., Arp 1966; Schweizer \& Seitzer 1988; Duc et al. 2014),  recent studies have detected much fainter analogues of these structures around spiral galaxies in the local universe, including the Milky Way (MW) and Andromeda. 
Moreover, recent extensive photometric databases have for the first time provided spectacular panoramic 
views of the MW tidal streams (Majewski et al. 2003; Belokurov et al. 2006; Slater et al. 2014) 
 and revealed the existence of large stellar sub-structures within the MW halo, 
 which have been interpreted as observational evidence to understand our Galaxy's past and present hierarchical formation. 
Likewise, the Pan-Andromeda Archaeological Survey (PAndAS; McConnachie et al. 2009) 
 has provided a panoramic view of the Andromeda halo,
 which reveals a striking diversity of halo sub-structures akin to those observed in the MW. 
These observations provide sound empirical support for the $\Lambda$-CDM prediction that tidally disrupted dwarf galaxies 
 are important contributors to stellar halo formation, at least for the Local Group spirals.  

While stellar streams in the MW and Andromeda can be studied in detail from studies using resolved stellar populations, 
 direct comparison of these features to cosmological models is complicated by `cosmic variance' ---
 the fact that each large spiral has experienced a unique hierarchical formation history that depends on
 the number and timing of satellite accretions, as well as both the internal kinematics
 and orbital properties of the individual satellites. 
A search for analogues to those features observed in the MW and Andromeda across a larger sample of nearby spiral galaxies 
 is required to understand if the recent merger histories for Local Group spirals are 'typical',  
 an issue that remains unclear (Mutch et al.~2011).  
Current $\Lambda$-CDM numerical simulations can guide this quest for star-stream observational signatures 
 (e.g., Johnston et al.~2008; Cooper et al.~2010). 
These simulations have demonstrated that the characteristics of sub-structure currently visible 
 in stellar halos are sensitive to recent (0-8 Gyr ago) merger histories of galaxies, 
 which is a timescale that corresponds to the last few tens of percent of mass accretion for a spiral galaxy like the MW. 
These models predict that a survey of $\sim$100 parent galaxies reaching 
 a surface brightness of $\sim$29 mag arcsec$^{-2}$ would reveal many tens of tidal features, 
 perhaps nearly one detectable stream per galaxy (at this surface brightness limit).
A direct comparison of these simulations with actual observations is not possible because no suitable dataset yet exists.
Thus, the observational portrait of satellite accretion events is far from complete. 

Recent deep, wide-field images of nearby MW analogue galaxies within the Local Volume 
 taken with small telescopes has revealed an assortment of 
 large-scale tidal structures in the halos of several nearby galaxies.
For the first time, the detected features exhibit the striking morphological diversity predicted by cosmological models  (Mart\'\i nez-Delgado et al.~2010; MD2010, hereafter). 
In addition, a systematic automated search around $\sim$450 spiral galaxies 
 in the Sloan Digital Sky Survey has provided the first estimate of the frequency of tidal features around spiral galaxies 
 to a limiting surface brightness of 27-28 mag arcsec$^{-2}$ ($\sim$ 10-15\%; Miskolczi et al. 2011).
Both observational projects have also yielded an unprecedented sample of bright stellar streams in nearby spiral galaxies, 
 including the discovery of observational analogues to the canonical morphologies predicted 
 from $N$-body models of stellar halos that are entirely constructed from satellite accretion (Johnston et al.~2008). In addition,
there are other ongoing projects to map the surroundings of nearby massive galaxies to detect tidal streams and new
satellites (e.g. Ludwig et al. 2012; Duc et al. 2014; van Dokkum \& Abrahan 2014).The detection of these structures offers a unique opportunity
 to study the still dramatic last stages of galaxy assembly 
in the local universe (e.g. Foster et al. 2014) 
 and to compare the frequency of tidal stellar features 
 against those predicted for MW-sized galaxies in the $\Lambda$-CDM paradigm. 

Encouraged from these results, the {\it Stellar Tidal Stream Survey} (PI.~Mart\'\i nez-Delgado) 
 is carrying out the first systematic survey of stellar tidal streams to a surface brightness sensitivity 
 of $\mu_{r}$=28.5 mag  arcsec$^{-2}$ using a network of small, robotic telescopes placed on different continents. 
In this paper, we report the discovery of a giant stellar stream in one of the classical and extensively studied  
 galaxies, NGC\,4631, also known as the ``Whale galaxy.'' 
Imaging observations and data reduction are described in Section \ref{sec:obs}. 
The stream is described and placed in the context of other sub-structure in NGC\,4631 
 in Section \ref{sec:stream}.
The stream properties are measured in Section \ref{sec:props} and used to construct
 a suitable $N$-body model in Section \ref{sec:model}.
The work is summarized and discussed in Section \ref{sec:conc}.

\section{OBSERVATIONS AND DATA REDUCTION }\label{sec:obs}

Deep imaging of the field around NGC\,4631 was collected at the Remote Observatory Southern Alps (ROSA, Verclause, France) 
 with a 40-cm aperture $f/3.75$ corrected Newtonian telescope.
A FLI ML16803 CCD camera was used and provided a pixel scale of 1.237$^{\prime\prime}$ px$^{-1}$
 over a 81$\arcmin \times 81 \arcmin$ field of view.
A set of individual 300 s images were obtained remotely with an 
 Astrodon Gen2 Tru-Balance E-series Luminance filter ($L$, hereafter) 
 over several photometric nights between December 2011 and February 2012.
As shown in Fig.~ 1$a$, this $L$ filter is a wide-band, nearly top-hat filter that transmits 
 from $380 \lesssim \lambda \; \mathrm{(nm)} \lesssim 680$, 
and broadly cover the $g$ and $r$ bands. 
Each individual exposure was reduced following standard image processing procedures for dark subtraction, bias correction, 
 and flat fielding adopted for the larger stream survey (MD2010).
The images were combined to create a final co-added image with a total exposure time of 23,100 s. 

With the aim of measuring the color of the detected structures (see Sec.~4), 
 we also construct $g$ and $r$ image mosaics from the  Sloan Digital Sky Survey (SDSS) 
 Data Release 7 (DR7; Abazajian et al.~2009) following the procedure described in Zibetti et al.~(2009).
Because these features are too faint to be detected visually in the SDSS mosaic at its native pixel scale, 
 these mosaics were resampled to the same pixel scale as our $L$ data (1.237$^{\prime\prime}$ px$^{-1}$).

The $L$ data were calibrated to SDSS DR7 $r$-band magnitudes as described by the following procedure.  
First, we remove any residual large-scale sky gradients from our wide-field images 
 by modeling the background with a two-dimensional, fourth-order Legendre polynomial.
The polynomial was fit to the median flux within coarse spatial bins that were 5$^{\prime}$ on a side 

after all $\geq5\sigma$ sources were masked. Aperture photometry was then performed on 27 isolated stars that were spatially
 distributed across the $L$ image, which were matched to the SDSS DR7 point source catalog (Abazajian et al.~2009).
Fig.~1$b$ compares the SDSS-measured $r$ magnitudes ($r_{\mathrm{SDSS}}$) 
 to the $L$ instrumental magnitudes ($L_{inst}$),  
 revealing a dependence on $g-r$ color spanning $\sim0.4$ mag over the full color range ($0.2 < \left(g-r\right) < 1.6$). 
The color residuals are expected due to the very different shapes of the $L$ and $r$-band filter transmission functions. 
Our flux calibration takes on the following form:
\begin{equation}
r_{\mathrm{cal}} = c_{0}\:L_{\mathrm{inst}} + c_{1} (g - r) + c_{2} \; ,
\end{equation}
where the three coefficients $c_{i}$ are determined from an iterative fit to the matched photometry for the 27 calibration stars. 
The final fit to the 27 SDSS calibration stars is given in Fig.~1$b$. 
In Fig.~1$c$, we show the residuals from our derived flux calibration relation for the 27 stars, 
 which have no remaining dependence on the $g-r$ color. 
The residuals have a standard deviation of 0.03 mag, which is adopted as the uncertainty in the flux calibration and is added in quadrature to the measurement errors reported hereafter.

\begin{figure*}[ht!]
\centering
\vspace{-4.5cm}
\includegraphics[width=14.5cm, angle=270]{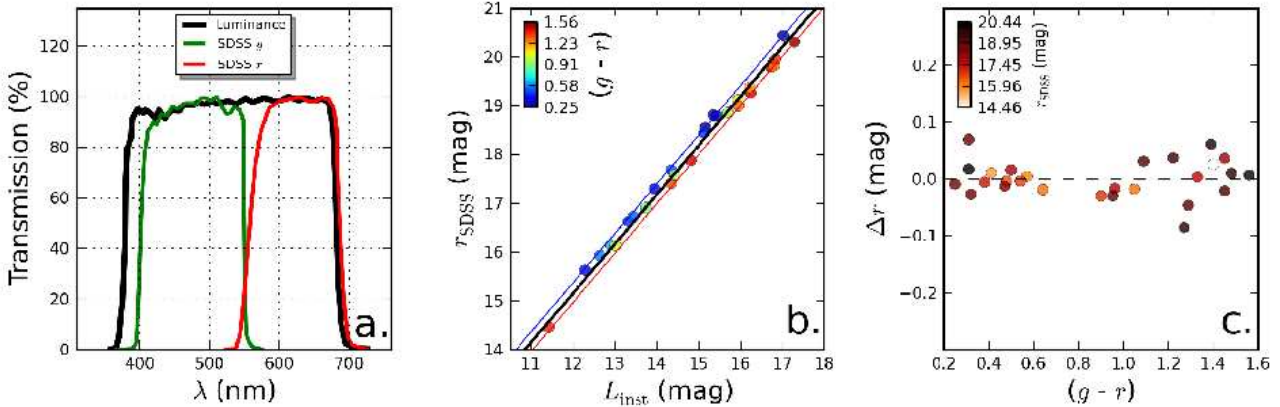}
\vspace{-4.5cm}
\caption{\label{fig:calib} 
 Calibration of the ROSA Luminance ($L$) photometry to the SDSS $r$-band. 
 {\bf (a)} Filter transmission curves for the wide-band $L$ filter compared to those of the SDSS $g$ and $r$-bands, 
 which are almost entirely contained {\it within} the pass-band of the $L$ filter. 
 {\bf (b)} The $r$ magnitudes for the 27 calibration stars in the NGC\,4631 field from the SDSS DR7 catalog ($r_{SDSS}$)
  compared to the corresponding instrumental $L$ magnitudes ($L_{\mathrm{inst}}$),
  with each point color coded by its $g-r$ color (as indicated in the inset color bar). 
  The three solid lines indicate the flux calibration relations 
   (Equation 1) evaluated at the bluest $g-r$ color (blue), 
   the median $g-r$ color (black), 
   and the reddest $g-r$ color (red) of the sample.
 {\bf (c)} The resulting magnitude residuals after application of the calibration relation 
    ($\Delta r = r_{\mathrm{SDSS}} - r_{\mathrm{L}}$), 
    compared to $g-r$ color with points color coded by $r_{\mathrm{SDSS}}$ (as indicated in the inset color bar). 
  The 1$\sigma$ spread in the residuals is 0.03 mag.
}
\end{figure*}

We use the method described in MD2010 for determining the limiting depth of our image, due to both photon noise and larger-scale background variations due to flat fielding, ghosts, and scattered light. To measure the photon noise limit, we calculate the mean of the standard deviation measured from 25 small $4\times4$ pixel boxes that were randomly distributed throughout the image, away from stellar and extended sources. To measure large-scale background fluctuations, we calculated five times the standard deviation of 25 randomly placed boxes in the background having a typical size comparable to the spatial extent of the faint features we intend to measure (here, $\sim1^{\prime}$ in size). Using $g-r = 0.5$, which is the midpoint of the range of $g-r$ colors of the features measured below, we find that the $5\sigma$ $r$-band limiting surface brightness of our $L$ data for photon noise is $25.8$ mag arcsec$^{-2}$ while the $5\sigma$ limit due to large-scale variations is $27.6$ mag arcsec$^{-2}$. In our measurements, we correct for foreground Galactic extinction by adopting the value appropriate at the coordinates of NGC\,4631  for the entire ROSA field of view ($A_{r} = 0.039$; Schlafly \& Finkbeiner 2011).

\section{The stellar stream of NGC 4631}\label{sec:stream}

Our deep $L$, wide field, panoramic view of the moderately interacting pair NGC\,4631/4656 (with a projected separation of $R_{ps}$ = 69 kpc) is given in Fig. 2. Our deep image reveals two low surface brightness, extended structures on either side of NGC\,4631.
We will refer to the `bridge-like' feature to the south-east of NGC\,4631 as stream$_{SE}$, and that to the north-west as stream$_{NW}$\footnote{This structure was also independently found by Karatchensev et al. (2014) in deep images obtained with amateur telescopes.}. 
The two features are roughly aligned with each other at a position angle (PA)  of $\sim140^{\circ}$,
 which suggests a common origin. 
If so, the discontinuity of the structure on the NW side of the galaxy could be due to the limiting surface brightness of the ROSA image. Interestingly, stream$_{NW}$ shows a more luminous inner region within the more elongated lower surface brightness feature. This structure is similar to that of a tidal disrupting dwarf galaxy (see for example NGC\,4216 in MD2010; Fig.~1 or Paudel et al. 2013) and suggests that it maybe the progenitor of stream$_{SE}$. We also note faint debris present between NGC\,4656 and its companion, NGC\,4656\,UV (Schechtman-Rook \& Hess 2012),
 indicative of interactions between these two galaxies. The detection of further, fainter structures is hampered by contamination from faint Galactic cirrus, which is barely visible in Fig.~2.

\begin{figure*}
\centering
\vspace{-4cm}
\includegraphics[width=18cm, angle=180]{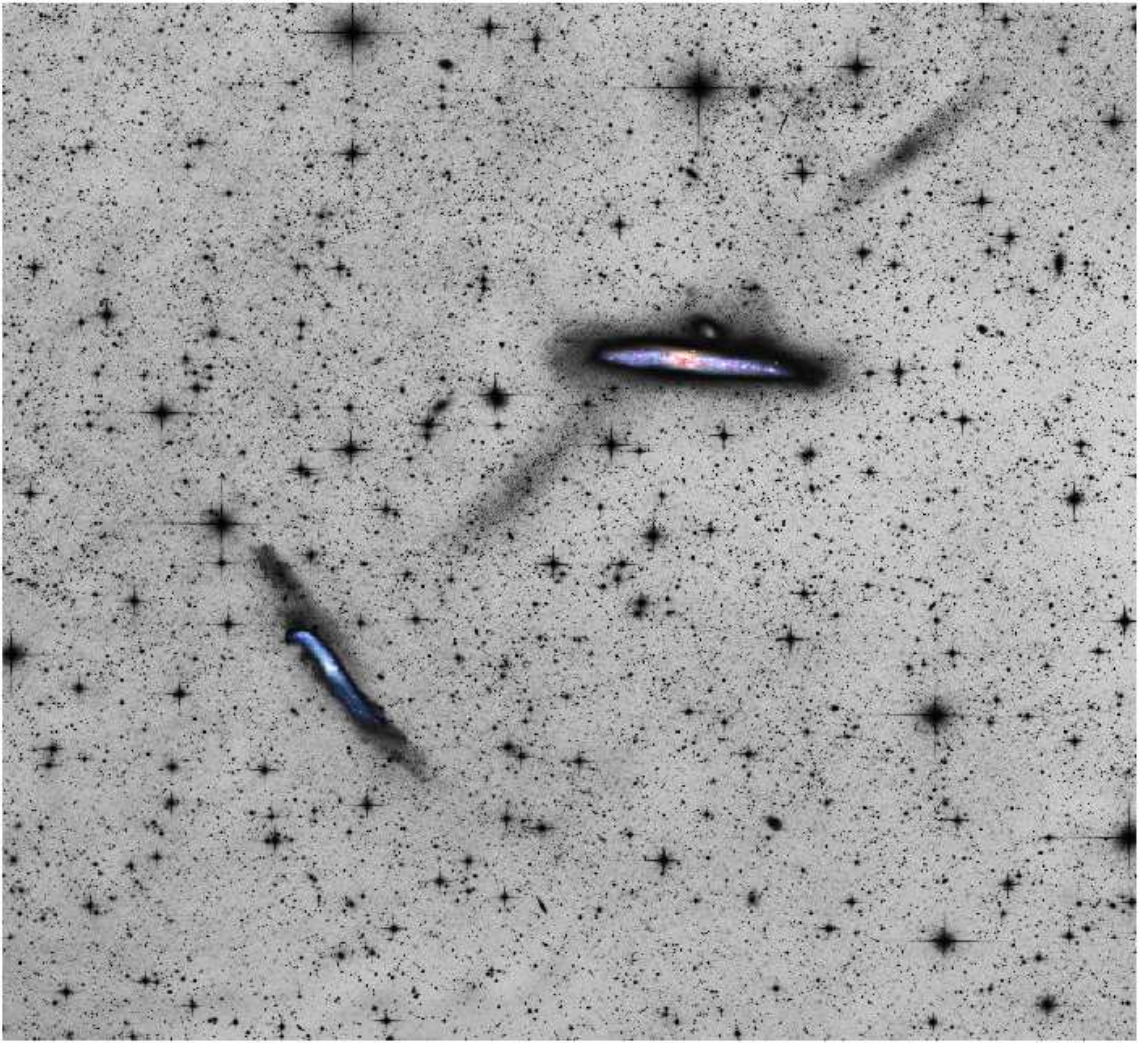}
\vspace{-4cm}
\caption{
 \label{fig:bigimage}
 Final ROSA $L$ image of the NGC\,4631 environs.
 The image is approximately 81$\arcmin$ $\times$ 81 $\arcmin$ in sky right orientation (e.g., East is to the left and North is up).
 NGC\,4631 (the 'Whale' galaxy) is the edge-on galaxy in the upper right and its large companion is NGC\,4656 to the lower left. 
 The projected separation of the two galaxies is $\sim$32$'$ or 69 kpc. 
 Several features are visible in the image, the most prominent being a stellar stream extending, in projection,
  from NGC\,4631 towards NGC\,4656 (stream$_{SE}$).
 Toward the North-West of NGC\,4631 is an overdensity with extensions along PA$=140^{\circ}$ (stream$_{NW}$), 
  whose extensions are roughly aligned with the PA of stream$_{SE}$.
 We also note the detection of debris between NGC\,4656 and its proposed tidal-dwarf companion, NGC\,4656\,UV.
}
\end{figure*}

Fig. 3 shows a zoomed region with an appropriate size to fit the NGC\,4631 stellar halo. For illustrative purpose,
we have included a color inset of the disk of the galaxy taken with the Black Bird Observatory during the pilot survey
of this project(MD 2010). This better resolves the vertical structures of the disk and the morphological disturbances of NGC\,4627, a dwarf Elliptical (dE) 
satellite of NGC\,4631 clearly interacting with its parent galaxy.
Although the extrapolation of the path of the stream$_{SE}$  across the disk passes near the position of NGC\,4631 in projection, 
the visible tidal arms of this companion are oriented almost perpendicular to the PA of the streams (see Fig.3).
Since streams trace the prints of their progenitor system, this suggests that NGC\,4627  is an unlikely candidate for the progenitor
 of stream$_{SE}$ and stream$_{NW}$.  It is more likely that NGC\,4627 is responsible for the optically 
 detected extended vertical structure within the NGC\,4631 disk (Ann et al.~2011)  or, given its overall blue mean color (Sec.~4), 
 to the complex gas streams described below.

\begin{figure}
\centering
\vspace{-2cm}
\includegraphics[width=8.5cm]{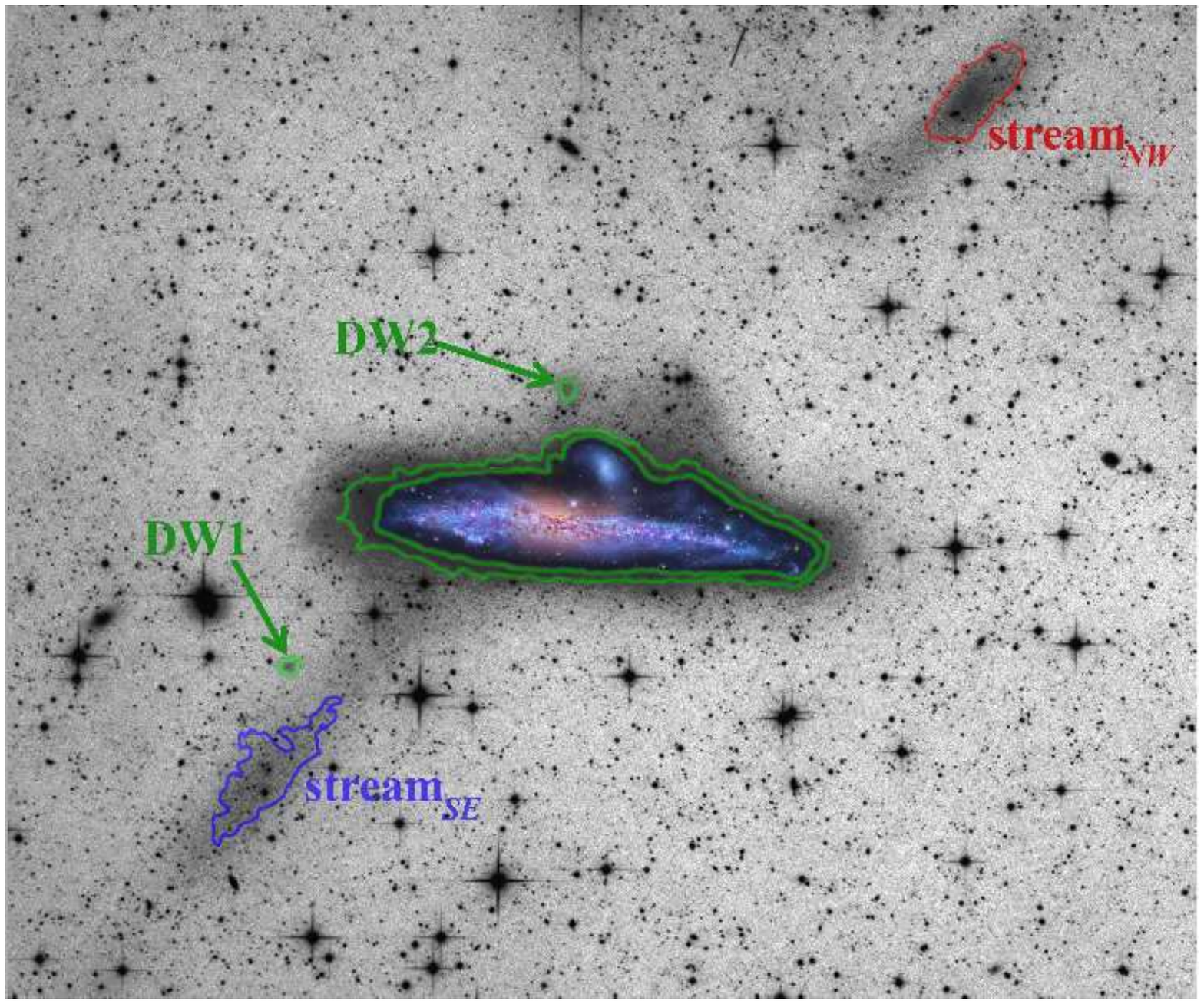}
\vspace{-2cm}
\caption{ 
 \label{fig:zoomimage}
 Zoom into the ROSA $L$ image (Fig.~1) to feature just the NGC\,4631 stellar halo 
  with isophotal boundaries overlaid for the major sub-substructures. 
 The image is approximately 42.5 $\arcmin$ $\times$ 34.6 $\arcmin$ in a sky right orientation.
 The two green isophotes around NGC4631 correspond to $\mu_{r} = 25$ mag arcsec$^{-2}$ and $\mu_{r} = 26$ mag arcsec$^{-2}$. 
 The small, green isophote around DW2 corresponds to $\mu_{r} = 26.5$ mag arcsec$^{-2}$, 
  which is the faintest closed isophote that excludes emission from the NGC\,4631 stellar halo. 
 The green isophote for DW1, the red isophote for stream$_{NW}$  and the blue isophote for the stream$_{SE}$
  all correspond to $\mu_{r} = 27.6\:(\pm0.1)$ mag/arcsec$^{-2}$. 
}
\label{overflow}
\end{figure}

An alternative explanation for the observed faint tidal features is that they  could be due to the
tidal tails from the interaction of NGC\,4631 with its large-spiral companion, NGC\,4656, a formation scenario that is
also proposed to explain a similar extended plume in NGC 3628 (Rots 1978).
However, as is discussed in Sec.~5, the edge-on perspective and visual configuration of both interacting systems and the morphology of the stream likely  rules out this scenario.

Assuming that the two stream-like pieces have a common origin 
 and are at the same distance of NGC\,4631 ($7.4\pm0.1$; Radburn-Smith et al.~2011), the total projected length of that structure is at least 85.2 kpc, as measured from the far edges of the ellipses fit to the isophotes of stream$_{SE}$ and the stream$_{NW}$, respectively. We also measured the widths of stream$_{SE}$ and the possible progenitor (within stream$_{NW}$)  by fitting ellipses to their isophotal boundaries (see Sec.~4). 
These ellipses are used only to determine the position angle of the isophote's major axis, 
 which we find to be the same within the uncertainties for each of the two faint features. 
We then collapse a strip of the image using a median combine (excluding pixels that are masked due to foreground stellar sources) 
 along the direction of the major axis. 
The measurement strip is centered on the minor axis of the isophote for each object, and is $\pm90^{\prime\prime}$ in width.
To measure the widths, we fit a Gaussian to each light profile, obtaining a FWHM of $4.69\pm0.07$ kpc and $2.53\pm0.04$ kpc for
stream$_{SE}$ and the progenitor candidate (within stream$_{NW}$), respectively. Stream$_{SE}$ is thus wider than stream$_{NW}$ by a factor of $\sim2$.

 Our deep image also detects the presence of two overdensities, marked as DW1 and DW2 in Fig.~3.
DW1, DW2, and the most luminous, inner portion of stream$_{NW}$ 
 were reported as dwarf companions of NGC\,4631 by Karatchensev et al.~(2014).
The first smaller object, DW1, is well separated from NGC\,4631 ($\sim$ 10$\arcmin$ South-East; 25 kpc) and it is clearly a faint dwarf spheroidal satellite of
 NGC\,4631.  The second DW2 feature is located  5$\arcmin$ ($\sim$ 12 kpc) North of the NGC\,4631 core 
 and its morphology is consistent with that of a faint, more extended dwarf galaxy satellite. Given the surface brightness limit of the $L$ image, however, we
 cannot rule out a connection to the vertical structure of the NGC\,4631.

In addition to the stellar streams and dwarf galaxy candidates already discussed, 
 the NGC\,4631 stellar halo contains a complex HI gas morphology.
In fact, it contains two relatively unique features:  
 (i) five individual HI tidal features extending from the disk (Rand 1994, Rand \& van der Hulst 1993);  
 (ii) two HI supershells associated with massive star forming regions that are expelling large masses of gas at
 high velocities (Rand \& van der Hulst 1993).  
Moreover, there is extensive diffuse H$\alpha$ emission throughout the gas halo (Donahue et al.~1995),
 spatially coincident with the gaseous streams.
We compare these data to investigate any potential associations between the stellar and gaseous substructures.

In Fig. 4, we overlay the low-resolution HI zeroth order map of Rand (1994) on the ROSA $L$ image.
The HI maps were obtained with the Very Large Array (VLA) at a resolution of 45$'' \times$ 89$''$ 
 over a velocity range of 110 to 1023 km s$^{-1}$ with velocity resolution of 10 km s$^{-1}$. 
The single-channel point source sensitivity is M$_{HI}$ =  7 $\times$ 10$^{5} M_{\odot}$
 for a source at the beam center.
We refer the reader to Rand (1994) for additional details on the observations and data processing. 
The five tidal spurs as identified in the Rand (1994) datasets are labelled.
Together, the spurs contain 3 $\times$ 10 $^{9}$ M$_{\odot}$ of HI gas,
 approximately one third of that measured for NGC\,4631 and half of that measured for NGC\,4656 (Rand 1994).
Note that the stellar stream identified in Fig.~1 is located between spurs 1 and 2.

\begin{figure}
\centering
\vspace{-2cm}
\includegraphics[width=8.5cm,angle=0]{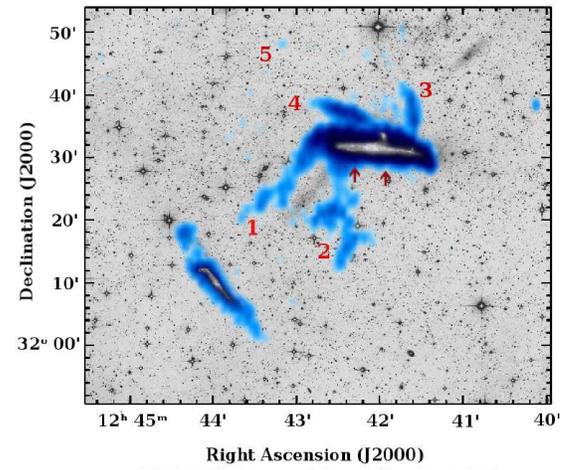}
\vspace{-2cm}
\caption{
\label{fig:h1map}
 Distribution of neutral hydrogen in the NGC\,4631 system (blue) overlaid on the ROSA $L$ image (Fig.~1)
  illustrating the complexity of the system. 
 The HI total intensities are taken from the VLA observations presented in Rand (1994)
  for the velocity range of 110 to 1023 km s$^{-1}$ 
 and is sensitive to a point source limit of 7 $\times$ 10$^{5}$ M$_{\odot}$. 
 The five HI tidal features ('spurs') are labeled (1-5) and the
  two HI super-shells are indicated with arrows.
 Stream$_{SE}$ is juxtaposed between spurs 1 and 2 and is near a region of enhanced H-$\alpha$ emission (Donahue et al.~1995).
 This orientation makes it unlikely that the HI features originated from the same satellite accretion event that 
  created Stream$_{SE}$.
 No HI is associated with stream$_{NW}$. 
 DW2 is embedded within Spur 4.}
\end{figure}

Tidal spurs 1 and 4 are thought to have been formed by interactions between NGC\,4631 
 and its large galaxy companion, NGC\,4656 (Combes 1978). However, it is unlikely to form stellar streams if 
the victim galaxy is the higher mass object in interaction (NGC 4656 
according to the observational data).
Spurs 2 and 3, on the other hand, were interpreted to be formed by tidal interactions between NGC\,4631 and NGC\,4627,
 its tidally distorted, blue dE companion.
Modelling of the tidal features, however, is complicated by the lack of reliable
 kinematic velocities for the dwarf companions in the field.

The projection of each of these features is interesting for many reasons. First, 
 the stellar stream discovered in this paper is located in between the shells 1 and 2.  
Second, the stream intersects the disk at one of the massive star forming regions. 
However, the stream itself is not of the morphology or optical colors 
 anticipated by the interaction of a gas rich galaxy 
 at the level needed to explain the gas streams (3$\times$10$^{9}$ M$_{\odot}$; Rand 1994). 
Moreover, the age of the stellar populations believed to drive HI supershells
 are far too young to have been affected by a satellite disk passage 
 (e.g., akin to the situation observed in NGC\,5387; Beaton et al.~2014).
Assuming a projected radius for the stream of $\sim$ 40-50 kpc, it would be 
 unlikely that the progenitor would pass in a gas-rich portion of the NGC\,4631 disk.
Thus, we suspect that neither the gas streams nor the star forming regions are associated with the stellar tidal stream.

\section{Photometric Properties and Stellar Mass Limits of the Stream}\label{sec:props}

Our analysis primarily focuses on the two most extended substructures in the NGC\,4631 stellar halo --  
 stream$_{NW}$ (that possibly includes the progenitor of the stream) and stream$_{SE}$ as identified in Fig.~3. 
Using the $L$ filter image, 
 we construct isophotes for each of the features corresponding to the $5\sigma$ large-scale background variation
 within the $L$ data. 
These isophotes are then overlaid on the SDSS mosaics for each feature 
 and used as boundaries inside of which the total integrated $g-r$ color 
 can be measured to a signal-to-noise ratio ($S/N$) that is sufficient for correcting the color term 
 required in Equation 1. 
In Fig.~3, these photometric boundaries are displayed for the features of interest, 
 which correspond to the $\sim27.6$ mag arcsec$^{-2}$ $r$-band surface brightness limit. 
Due to $S/N$ limitations of our data, 
 there may be faint emission associated with these features that lies outside of these boundaries. Therefore, 
 the luminosities we report are lower limits.

We proceed to measure the integrated $g-r$ color of stream$_{NW}$ and the stream$_{SE}$ from the 
 SDSS mosaics, which have had all foreground sources masked.
Each $g$ and $r$ mosaic is corrected for foreground Galactic extinction
 by adopting $A_{g} = 0.056$ and $A_{r} = 0.039$ at the position of NGC\,4631
 for the entire field of view (Schlafly \& Finkbeiner 2011).
The stream$_{NW}$ feature has $g-r = 0.32 \pm 0.02$, which is relatively blue. 
The stream$_{SE}$ is significantly redder at $g-r = 0.69 \pm 0.02$. 
We also measured the $g-r$ color of over-densities DW1 and DW2 (see Fig. 3 and Sec.~3). 
For DW1, we measure $g-r = 0.59 \pm 0.08$. 
Since DW2 is located spatially near the inner stellar halo of NGC\,4631, 
 we measure the $g-r$ color inside of the faintest isophote that closes without including emission from the larger galaxy. 
This isophote corresponds to $\mu_{r} = 26.5$ mag arcsec$^{-2}$, 
 the $g-r$ color inside of which is $0.79\pm0.07$. 
While stream$_{SE}$ has a $g-r$ color comparable to DW1 and DW2, 
 the color of stream$_{NW}$ is significantly bluer. 
For comparison, the integrated $g-r$ color of the NGC4631 stellar halo between 
 $25 < \mu_{r}$ (mag arcsec$^{-2}$) $< 26$ is $g-r = 0.23\pm0.01$. 
The color difference between stream$_{SE}$ and stream$_{NW}$ suggests that the stellar populations 
 in stream$_{NW}$ could be on average younger --- 
 an observation that largely supports the hypothesis that stream$_{NW}$ contains the stream progenitor. 

Using the integrated $g-r$ colors measured for the extended faint features listed above, 
 we can correct for the color term in the $L$-filter image calibration to obtain high $S/N$ 
 average surface brightness measurements over the spatial extent of the objects. 
For stream$_{NW}$, we measure $\left\langle \mu_{r} \right\rangle = 25.92\pm 0.05$ mag arcsec$^{-2}$. 
The stream$_{SE}$ is fainter with $\left\langle \mu_{r} \right\rangle = 26.16\pm 0.05$ mag arcsec$^{-2}$. 
 To check our measurements, we have also calculated the surface brightness within these same photometric boundaries directly
 from the SDSS $r$-band mosaic. We find that the difference between the surface brightness measured from the $L$-filter image
 and that from the SDSS $r$-band mosaic is consistent with zero for all features measured here within $2\sigma$, which corresponds to 0.14 mags. 
The photometric properties measured for stream$_{NW}$ and the stream$_{SE}$ as described in this section are tabulated in Table 1.

\begin{deluxetable}{lccc}
\tabletypesize{\small}
\tablecaption{Surface Photometry of Sub-Structure around NGC\,4631}
\tablehead{
  \colhead{} & \colhead{} & \colhead{stream$_{NW}$} & \colhead{stream$_{SE}$} }
\startdata
$g-r$                                & (mag)               & $0.32\pm0.02$       & $0.69\pm0.02$    \\
$\left\langle \mu_{r} \right\rangle$ & (mag arcsec$^{-2}$) & $25.92\pm 0.05$     & $26.16\pm 0.05$    \\
FWHM                                 & (kpc)               & $2.53\pm0.04$       & $4.76\pm0.07$     \\
L$_{r}$                              & (L$_{\odot,r}$)     & $>1.61\times10^{7}$ & $>1.45\times10^{7}$   \\
M$_{*}$                              & (M$_{\odot}$)       & $>7.2\times10^{6}$  & $>2.6\times10^{7}$    \\
\enddata
\end{deluxetable}

To estimate a stellar mass for stream$_{NW}$ and stream$_{SE}$, 
 we must obtain a measurement of the total luminosity ($L_{r}$) from our $r$-band calibrated $L$ image 
 and an estimate of the stellar mass-to-light ratio $\Upsilon_{r}$. 
To calculate the former, we measure the total flux within the isophotal boundaries. 
Note that pixels within the boundary that are masked for foreground stellar sources 
 are replaced by averaging linearly interpolated values along pixel rows and columns. 
Taking into account the adopted distance to NGC\,4631 
 and recalling that additional faint emission may lie outside of the isophotal boundaries, 
 the measured fluxes correspond to lower limits on the $r$-band stellar luminosity 
 of $1.61\times10^{7}$ L$_{\odot,r}$ for stream$_{NW}$ 
 and $1.45\times10^{7}$ L$_{\odot,r}$ for the stream$_{SE}$. 
We utilize the relations giving $\Upsilon_{r}$ as a function of various color indexes from 
 Zibetti et al.~(2009), who assume a Chabrier initial mass function (Chabrier 2003). 
For the $g-r$ color index, the relation is:
\begin{equation}
\Upsilon_{r} (g-r) = 10^{-0.840 + 1.654 (g-r)}
\end{equation}
 which yields $\Upsilon_{r} = 0.49 \pm 0.04$ for stream$_{NW}$ 
 and $\Upsilon_{r} = 2.0 \pm 0.2$ for stream$_{SE}$. 
Using these values, we estimate a stellar mass lower limit M$_{*}$ of $7.4\times10^{6}$ M$_{\odot}$ for stream$_{NW}$ 
 and $2.7\times10^{7}$ M$_{\odot}$ for stream$_{SE}$. 
These values are also tabulated in Table 1. If considering that both features are part of the same large tidal structure, a lower limit on its total stellar mass would be $3.3\times10^{7}$ M$_{\odot}$. 

\section{A model for the stellar stream} \label{sec:model}

The aim of this section is to explore whether the observed stream-like feature around NGC\,4631 
 can be directly linked to interactions between this spiral galaxy and one of its dwarf companions by means of a theoretical model.
We also note, as previously demonstrated by Toomre \& Toomre (1972) and D'Onghia et al.~(2010),
that tidal tails and bridges connecting galaxy pairs continue both to lengthen and to thin out after the encounter.
It was shown in these studies that the efficiency of tail formation depends on the inclinations      
 of the disks relative to the orbital plane.
NGC\,4656 and NGC\.4631 are both almost edge-on (e.g. see Fig.1) and, in particular,
 NGC\,4656 is inclined almost 90 degrees relative to the orbital plane of NGC\,4631.
This configuration inhibits the formation of any tail or stellar stream and excludes that
 any effective tidal interaction between NGC\,4656 and NGC\,4631 is responsible of the observed extended structure.

In our numerical experiment, we only chose to focus on the extended stellar stream around NGC\,4631. 
Thus, the goal is not to reproduce every detail of this complex system ---
 the interactions of the numerous companions, the HI streams, and the stellar streams --
 but rather to produce a viable model for the feature discovered here.
A model incorporating all of the observed debris features in the NGC\,4631 halo
 and its interaction with NGC\,4565 would require a parameter search of all the possible orbital configurations and mass ratios
 for each of the galaxies,
 which is beyond the scope of the work presented here. 

We follow the general method outlined in D'Onghia et al.~(2009) to set up the
 initial galaxy models and satellite orbits 
 to reproduce the observed large-scale stellar stream around NGC\,4631. 
In the model, NGC\,4631 consists of a dark matter halo and a rotationally supported stellar disk. 
The parameters describing each component are independent 
 and the models are constructed in a manner similar to the approach presented in previous works.
We model the dark matter mass distribution of each galaxy with a Hernquist profile (1990) and an exponential stellar disk.  
The total halo mass of NGC\,4631 is assumed to be M$_{halo}$ = 4x10$^{11}$ M$_{\odot}$ and a halo scale length of 22 kpc. 
The disk is a particle-realization of an exponential disk with a constant scale-height,
 in which both the positions and velocities of the particles are assigned following the prescriptions in Springel et al.~(2005). 
We adopt a radial scale length for the disk of R$_{disk}=$2 kpc, with scale-height given by z$_{0}$ = 0.1 R$_{disk}$. 
The rotation curve of NGC\,4631 peaks at 130 km$s^{-1}$ in agreement with observational estimates (Combes 1977, and references therein). 
 
We model a dwarf satellite galaxy consisting of a dark matter halo of  10$^{10}$ M$_{\odot}$, 
 a scale length of 9 kpc and a stellar mass fraction of 4\% of the total mass.
The number of particles of each component (N$_{*}$ for the number of star particles stars,
 N$_{halo}$ for dark matter particles) is chosen such that the mass
 resolution per particle of a given type is roughly the same for both galaxies.
We adopt $N_{halo}$=10$^5$;10$^4$ and N$_{*}$=10$^6$;10$^{5}$ to model NGC\,4631 and its orbiting dwarf, respectively.

Generally, higher orbital eccentricities for the satellite result in `fly-by' encounters,
 while lower values cause the dwarf's orbit to decay rapidly.
Given the morphology of the debris, 
 the dwarf galaxy is placed on an eccentric orbit about NGC\,4631 (e = 0.6).
With this given orbit the pericentric distance of the dwarf from the center of NGC\,4631 is around $R_{peri}=$15 kpc
 and the apocentric distance is $R_{apo}=$60 kpc.

The satellite is placed in an orbit around NGC\,4631 and is evolved over a period of $\sim$3.5 Gyr. 
The simulations were carried out with GADGET3, a parallel TreePM-Smoothed particle
 hydrodynamics (SPH) code developed to compute the evolution of stars and dark matter, 
 both of which are treated as collisionless ﬂuids.

Fig.~4 displays the resulting $N$-body model 
 showing the density of star particles at t$\sim$ 3.5 Gyr. 
During the simulations,
 stars are pulled out of the dwarf satellite in its orbit around NGC\,4631 to form a long stream 
 that is in qualitative agreement with that observed stream displayed in Fig.~2 and Fig.~3.
In the model, what remains of the dwarf is contained within the simulated analog of the 
 stream$_{NW}$ feature. Assuming that our derived stream stellar mass lower limit ($3.3\times10^{7}$ M$_{\odot}$; see
Sec. 4) is $\sim$20\% of the total mass of the progenitor, as shown by simulations of merging galaxies (Cox et al. 2006), our simulations suggests a total initial stellar mass of $3.3\times10^{8}$ M$_{\odot}$ for
the dwarf  and a total dynamical mass (including dark matter) of $\sim 10^{10}$ M$_{\odot}$.
Thus the mass ratio of the interaction between NGC4631 and the dwarf satellite causing the observed stellar stream is approximately 40.

\begin{figure*}
\centering
\vspace{-2.5cm}
\includegraphics[width=12cm, angle=0]{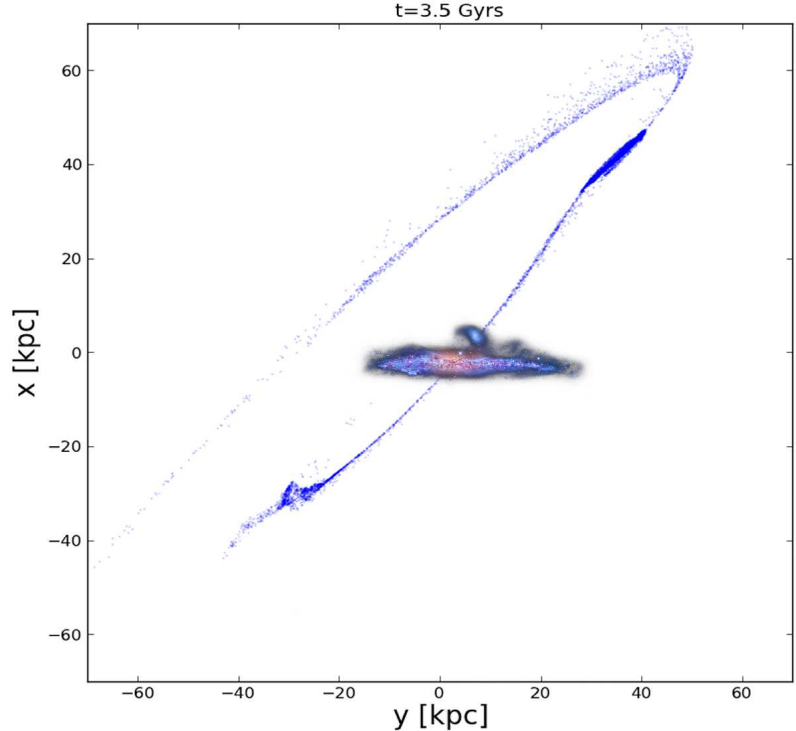}
\vspace{-2.5cm}
\caption{
 \label{fig:model}
 Snapshot of the $N$-body simulation that qualitatively reproduces both stream features in NGC\,4631
  from an interaction between it and a single dwarf satellite on a mildly eccentric orbit (e=0.6). The stream is displayed at t$\sim 3.5$ Gyr into the interaction.
 Only a single interaction, that creating the stellar stream, is modelled. 
 Owing to the numerous dwarf companions and tidal debris observed in NGC\,4631,
  a far more complex model would be required to reproduce 
  all of the features and is beyond the scope of this work.
}
\end{figure*}

\section{Conclusions} \label{sec:conc}

We report the discovery of an extended stellar tidal stream around NGC\,4631. 
From our data, combined with arguments based on simulations of tidal interactions,
 we conclude that this newly detected stellar stream cannot be due to interactions between NGC\,4631 and the galaxy NGC\,4656.
Intriguingly, the discovered stellar stream lies between previously detected HI streams (referred to as spurs 1 and 2),
 which indicates that the stellar stream is most likely not related to these features and thereby produced by a different progenitor satellite.
 This stellar substructure is added to an already rich census of substructure in the NGC\,4631 halo,
 including five HI `spurs' with overall structure very similar to that of the Magellanic Stream in the MW halo 
 and in two HI supershells associated with massive star forming regions in its disk.  Thus, the NGC\,4631 stellar halo contains numerous on-going
 accretion events, akin to the complexity observed in the stellar halos of the Local Group spirals. 

The feature observed around NGC\,4631 is reproduced by numerical simulations of the tidal interaction
 of a candidate dwarf galaxy (embedded in stream$_{NW}$)  with its parent, NGC\,4631.
The model qualitatively  agrees with our assessment from the optical morphology of the features.

We recovered two previously detected faint dwarf galaxy candidates  associated with the NGC\,4631 system that could be the progenitors of the
 gas streams. For example, the DW1 might be associated to spur 1 and a possible candidate DW2 could be associated with spur 4. 
However, the investigation of the full complement of dwarf satellite interactions and debris features within the NGC\,4631 halo is beyond the scope of this paper and it will be explored in a forthcoming study.

A visual inspection of the projected positions of dwarf candidates around NGC 4631  and the path of the
stellar stream might suggest that the substructures might be alligned
within an extended plane. Similar configurations have been claimed
to be found among satellite galaxies around M31 (Koch\& Grebel 2006; Ibata et al 2014) and our Milky Way
(Palowski et al. 2012). It has been suggested that this confinement of satellite galaxies in thin planes 
around galaxies presents a challenge to the theory of hierarchical galaxy clustering.
(Palowski et al. 2014; D'Onghia \& Lake 2008). In this context, NGC4631 and its complex system of
 satellite galaxies represent an additional interesting case to study the
problem of how these satellites fell into a more massive system. However, follow-up
observations to derive accurate distances (by means of its resolved stellar population) and radial velocities are needed
 to shed light on the actual spatial configuration of these structures and their possible common origin.

\acknowledgements{We thank R. Rand for the use of his VLA HI data cube
 and M. Donahue for a physical copy of her 1995 paper for comparison to our new image.
ED gratefully acknowledges the support of the Alfred P. Sloan Foundation. ED and RLB express their appreciation towards the Aspen Center for Physics and the NSF Grant N. PHYS-1066293 for hospitality during the writing of this paper. T.S.C. acknowledges the support of the National Science Foundation Graduate Research Fellowship. SDSS science image mosaics are produced using the package \textit{SDSSmosaic} developed by Stefano Zibetti. Funding for the SDSS and SDSS-II has been provided by the Alfred P. Sloan Foundation, the Participating Institutions, the National Science Foundation, the U.S. Department of Energy, the National Aeronautics and Space Administration, the Japanese Monbukagakusho, the Max Planck Society, and the Higher Education Funding Council for England. The SDSS Web Site is \textit{http://www.sdss.org/}. The SDSS is managed by the Astrophysical Research Consortium for the Participating Institutions. The Participating Institutions are the American Museum of Natural History, Astrophysical Institute Potsdam, University of Basel, University of Cambridge, Case Western Reserve University, University of Chicago, Drexel University, Fermilab, the Institute for Advanced Study, the Japan Participation Group, Johns Hopkins University, the Joint Institute for Nuclear Astrophysics, the Kavli Institute for Particle Astrophysics and Cosmology, the Korean Scientist Group, the Chinese Academy of Sciences (LAMOST), Los Alamos National Laboratory, the Max-Planck-Institute for Astronomy (MPIA), the Max-Planck-Institute for Astrophysics (MPA), New Mexico State University, Ohio State University, University of Pittsburgh, University of Portsmouth, Princeton University, the United States Naval Observatory, and the University of Washington. }




{}

\end{document}